\documentclass[conference,twocolumn]{IEEEtran}

\usepackage{graphicx}
\usepackage{color}
\usepackage{placeins}
\usepackage{float}
\usepackage{tabularx,colortbl}
\usepackage{amssymb}
\usepackage{amsthm}
\usepackage{cite}
\usepackage{amsmath}

\theoremstyle{plain}
\newtheorem{thm}{Theorem}

\theoremstyle{plain}
\newtheorem{rem}{Remark}
\newtheorem{cor}{Corollary}

\IEEEoverridecommandlockouts

\begin{document}

\title{Cell-Free Massive MIMO with Channel Aging and Pilot Contamination}

\author{Jiakang Zheng, Jiayi~Zhang, Emil~Bj\"{o}rnson, and Bo Ai
\thanks{J. Zheng and J. Zhang are with the School of Electronics and Information Engineering, Beijing Jiaotong University, Beijing 100044, P. R. China.}
\thanks{E. Bj\"{o}rnson is with the Department of Electrical Engineering (ISY), Link\"{o}ping University, Link\"{o}ping, Sweden. (e-mail: emil.bjornson@liu.se).}
\thanks{B. Ai is with the State Key Laboratory of Rail Traffic Control and Safety, Beijing Jiaotong University, Beijing 100044, China.}
}

\maketitle

\begin{abstract}
In this paper, we investigate the impact of channel aging on the performance of cell-free (CF) massive multiple-input multiple-output (MIMO) systems with pilot contamination. To take into account the channel aging effect due to user mobility, we first compute a channel estimate. We use it to derive novel closed-form expressions for the uplink spectral efficiency (SE) of CF massive MIMO systems with large-scale fading decoding and matched-filter receiver cooperation.
The performance of a small-cell system is derived for comparison. It is found that CF massive MIMO systems achieve higher  95\%-likely uplink SE in both low- and high-mobility conditions, and CF massive MIMO is more robust to channel aging.
Fractional power control (FPC) is considered to compensate to limit the inter-user interference. The results show that, compared with full power transmission, the benefits of FPC are gradually weakened as the channel aging grows stronger. 
\end{abstract}

\IEEEpeerreviewmaketitle

\section{Introduction}
Cell-free (CF) massive multiple-input multiple-output (MIMO) has been recently proposed as a future technology for providing a more uniform spectral efficiency (SE) to the user equipments (UEs) in wireless networks \cite{zhang2019multiple,Ngo2017Cell}. CF massive MIMO systems consist of many geographically distributed access points (APs) connected to a central processing unit (CPU) for coherently serving the UEs by spatial multiplexing on the same time-frequency resource \cite{Ngo2017Cell}. The characteristic feature of CF massive MIMO, compared with traditional cellular systems, is the operating regime with no cell boundaries and many more APs than UEs \cite{bjornson2019making}. In conventional small-cell systems, the APs only serve UEs within their own cell, which may lead to high inter-cell interference. Results in \cite{Ngo2017Cell} and \cite{bjornson2019making} show that CF massive MIMO systems outperform small-cell systems in terms of 95\%-likely per-user uplink and downlink SE.
Following these seminal works, many important and fundamental aspects of CF massive MIMO have been studied in recent years. For example, the authors in \cite{nayebi2017precoding} observed that CF massive MIMO systems with the large-scale fading decoding (LSFD) receiver achieve two-fold gains over the matched filter (MF) receiver in terms of 95\%-likely per-user SE.
The key difference between LSFD and MF receivers is that the former requires more statistical parameters to be available at \mbox{the CPU \cite{bjornson2019making,zheng2020efficient}}.

Most of the current works on CF massive MIMO consider a block-fading model, i.e., the channel realization in a coherence block is approximated as constant \cite{Ngo2017Cell,bjornson2019making,nayebi2017precoding,zheng2020efficient}.
However, practical channels are continuously evolving due to UE mobility, leading to the so-called channel aging effect where the channel is different but correlated between samples in a transmission block \cite{truong2013effects}.
The impact of channel aging has been characterized in co-located massive MIMO systems. For instance, the authors in \cite{yuan2020machine} utilized machine learning to predict the channels in massive MIMO systems under channel aging. The performance of massive MIMO systems under channel aging was also investigated in \cite{chopra2018performance,papazafeiropoulos2016impact}.
To the best of our knowledge, this is the first time that the channel aging effect is analyzed for CF massive MIMO systems.

Motivated by the above observations, we investigate the \mbox{uplink} performance of CF massive MIMO systems with LSFD and MF receivers under channel aging and pilot contamination. The performance of the corresponding small-cell system is analyzed for comparison. We derive a channel estimator and compute closed-form expressions for the uplink SE of the considered systems. Our results show that the CF massive MIMO system performs better than the small-cell system in both static and mobile scenarios. In particular, CF massive MIMO is more robust to channel aging. Finally, a practical fractional power control (FPC) is applied to further improve the system performance.

\textbf{Notation:} Column vectors and matrices are represented by boldface lowercase letters $\mathbf{x}$ and boldface uppercase letters $\mathbf{X}$, respectively. We use superscripts $x^\mathrm{*}$ and $\mathbf{x}^\mathrm{H}$ to \mbox{represent} conjugate and conjugate transpose, respectively.
We use ${\mathrm{diag}}\left( {{x_1}, \ldots ,{x_n}} \right)$ for a block-diagonal matrix with the \mbox{variables} ${{x_1}, \ldots ,{x_n}}$ on the diagonal.
The Euclidean norm, the expectation operators and the definitions are denoted by $\left\|  \cdot  \right\|$, $\mathbb{E}\left\{  \cdot  \right\}$, and $\triangleq$, respectively.
Finally, $x \sim \mathcal{C}\mathcal{N}\left( {0,{\sigma^2}} \right)$ represents a circularly symmetric complex Gaussian random variable (RV) $x$ with variance $\sigma^2$.

\vspace{-2mm}

\section{System Model}\label{se:model}

We consider a CF massive MIMO system consisting of $L$ APs and $K$ mobile UEs.
Each AP and UE is equipped with a single antenna. The APs are connected to a CPU via fronthaul links. We assume that all $L$ APs simultaneously serve all $K$ UEs on the same time-frequency resource. The UEs are assumed to move at different speeds, which affect the channel variations.
The communication is divided into resource blocks consisting of $\tau_c$ time instants (channel uses).\footnote{Note that a resource block is not the same as a coherence block in the block fading model often considered in the literature \cite{bjornson2017massive}. We consider a more realistic model where the channels are continuously evolving in a resource block and correlated between channel uses.}
We assume that the uplink training phase occupies $\tau_p$ time instants and the remaining $\tau_c-\tau_p$ time instants are used for payload data.
At the $n$th time instant in a given block, the Rayleigh fading channel between AP $l$ and UE $k$ is modelled as
\begin{align}
{h_{kl}}\left[ n \right] \sim {\mathcal{CN}}\left( {0,{\beta _{kl}}} \right),\ n =0,1, \ldots ,{\tau _c},
\end{align}
where $\beta _{kl}$ is the large-scale fading coefficient. Note that ${h_{kl}}\left[ n \right]$ is independent for different pairs of $k=1, \ldots ,K$ and $l=1, \ldots ,L$. However, ${h_{kl}}\left[ 0 \right], \ldots, {h_{kl}}\left[ \tau_c \right]$ are correlated.

\subsection{Channel Aging}
The relative movement between the UEs and APs lead to temporal variations in the propagation environment which affect the channel coefficient also within a resource block.
The channel realization ${h_{kl}}\left[ {n } \right]$ can be modeled as a function of its initial state ${h_{kl}}\left[ 0 \right]$ and an innovation component \cite{chopra2018performance}, such as
\begin{align}\label{channelaging}
{h_{kl}}\left[ {n } \right] = {\rho _k}\left[ {n } \right]{h_{kl}}\left[ 0 \right] + {\bar \rho_k }\left[ {n } \right]{g_{kl}}\left[ {n} \right],
\end{align}
where ${g_{kl}}\left[ {n } \right] \sim \mathcal{C}\mathcal{N}\left( {0,{\beta _{kl}}} \right)$ represents the independent innovation component at the time instant $n$.
In addition, $\rho_k\left[ {n} \right]$ represents the temporal correlation coefficient of UE $k$ between the channel realizations at time $0$ and $n$, and ${{\bar \rho }_k}\left[ n \right] = \sqrt {1 - {\rho^2_k}\left[ n \right]} $.
As in \cite{chopra2018performance}, we consider ${\rho _k}\left[ n \right] = {J_0}\left( {2\pi {f_{D,k}}{T_s}n} \right)$, where ${J_0}\left(  \cdot  \right)$ is the zeroth-order Bessel function of the first kind \cite[Eq. (9.1.18)]{Abramowitz1964table}, $T_s$ denotes the sampling time, and $f_{D,k}=(v_k f_c)/c$ is the Doppler shift for a UE with velocity $v_k$, where $f_c$ and $c$ represent the carrier frequency and the speed of light, respectively.
\begin{rem}
The model in \eqref{channelaging} is not a first-order autoregressive model, as considered in previous works such as \cite{truong2013effects}, but the correlation coefficients are selected to approximately match the Jakes' model \cite{chopra2018performance}.
In practice, the model \eqref{channelaging} is only accurate during a limited period of time since wide-sense stationarity can only be guaranteed for small-scale movements. To avoid this issue, we are not exploiting correlation between blocks, even if it might exist.
\end{rem}

\begin{figure*}[t!]
\setcounter{equation}{9}
\begin{align}
  &{{\hat s}_k}\!\left[ n \right] \!= \!\!\sum\limits_{l = 1}^L \!{a_{kl}^ *\! [n]{{\overset{\lower0.5em\hbox{$\smash{\scriptscriptstyle\smile}$}}{s} }_{kl}}}\! \left[ n \right] \!=\! \underbrace {{\rho _k}\!\left[ {n\! -\! \lambda } \right]\!\sum\limits_{l = 1}^L \!{a_{kl}^ *\![n] \mathbb{E}\!\left\{\! {\hat h_{kl}^* \!\left[\lambda  \right]\!{h_{kl}}}\!\left[\lambda  \right] \!\right\}} \!{s_k}\!\left[ n \right]}_{{\mathrm{D}}{{\mathrm{S}}_{k,n}}} \!+ \!\underbrace {{\rho _k}\!\left[ {n \!-\! \lambda } \right]\!\!\left( {\sum\limits_{l = 1}^L \!{a_{kl}^ *\! [n] \!\left(\! {\hat h_{kl}^*\! \left[\lambda  \right]\!{h_{kl}} \!\left[\lambda  \right] \!-\! \mathbb{E}\!\left\{ \!{\hat h_{kl}^* \!\left[\lambda  \right]\!{h_{kl}}} \!\left[\lambda  \right] \!\right\}} \!\right)} } \!\right)\!{s_k}\!\left[ n \right]}_{{\mathrm{B}}{{\mathrm{U}}_{k,n}}} \notag \\
   &+ \underbrace {{{\bar \rho }_k}\left[ {n - \lambda } \right]\sum\limits_{l = 1}^L {a_{kl}^ *[n] \hat h_{kl}^* \left[\lambda \right]{\vartheta_{kl}}\left[ n \right]} {s_k}\left[ n \right]}_{{\mathrm{C}}{{\mathrm{A}}_{k,n}}} + \sum\limits_{i \ne k}^K {\underbrace {\sum\limits_{l = 1}^L {a_{kl}^ * [n] \hat h_{kl}^* \left[\lambda  \right]{h_{il}}\left[ n \right]{s_i}\left[ n \right]} }_{{\mathrm{U}}{{\mathrm{I}}_{ki,n}}}}  + \underbrace {\sum\limits_{l = 1}^L {a_{kl}^ * [n] \hat h_{kl}^* \left[\lambda  \right]{w_l}\left[ n \right]} }_{{\mathrm{N}}{{\mathrm{S}}_{k,n}}}.  \label{uatf}
\end{align}
\hrule
\end{figure*}

\newcounter{mytempeqncnt}
\begin{figure*}[t!]
\setcounter{mytempeqncnt}{1}
\setcounter{equation}{10}
\begin{align}\label{SINR_k}
{\mathrm{SE}}_k^{{\mathrm{CF}}} = \frac{1}{{{\tau _c}}}\sum\limits_{n = \lambda }^{{\tau _c}} {{{\log }_2}\left( {1 + \frac{{\rho _k^2\left[ {n - \lambda } \right]{p_k}{{\left| {{\mathbf{a}}_k^{\mathrm{H}}\left[ n \right]{{\mathbf{b}}_k}} \right|}^2}}}{{\sum\limits_{i = 1}^K {{p_i}{\mathbf{a}}_k^{\mathrm{H}}\left[ n \right]{{\mathbf{\Gamma }}_{ki}}{{\mathbf{a}}_k}\left[ n \right]}  + \sum\limits_{i \in {\mathcal{P}_k}\setminus \{ k \}} {\rho _i^2\left[ {n - \lambda } \right]{p_i}{{\left| {{\mathbf{a}}_k^{\mathrm{H}}\left[ n \right]{{\mathbf{c}}_{ki}}} \right|}^2}}  + {\sigma^2}{\mathbf{a}}_k^{\mathrm{H}}\left[ n \right]{{\mathbf{\Lambda }}_k}{{\mathbf{a}}_k}\left[ n \right]}}} \right)}.
\end{align}
\hrulefill
\setcounter{equation}{2}
\end{figure*}

\subsection{Uplink Channel Estimation}

We assume that $\tau_p$ mutually orthogonal time-multiplexed pilot sequences are utilized. This means that pilot sequence $t$ corresponds to sending a pilot signal only at time instant $t$. This pilot design is necessary to keep the orthogonality in the presence of channel aging.
We consider a large network with $K>\tau_p$ so that different UEs will be assigned to the same time instant.
The index of the time instant assigned to UE $k$ is denoted by ${t_k} \in \left\{ {1, \ldots ,{\tau _p}} \right\}$ and the other UEs that use the same time instant for pilot transmission as UE $k$ is denoted by ${\mathcal{P}_k}= \{ i : t_i = t_k \} \subset \left\{ {1, \ldots ,K} \right\}$.
The received signal between AP $l$ and UE $k$ at time instant $t_k$ is
\begin{align}\label{zl}
{z_l}\left[ t_k \right] = \sum\limits_{i \in {\mathcal{P}_k}} {\sqrt {{p_i}} {h_{il}}\left[ t_i \right]} + {w_l}\left[ t_k \right],
\end{align}
where ${p_i} \geqslant 0$ is the pilot transmit power of UE $i$ and ${{w}_l}\left[ t_k \right] \sim {\mathcal{CN}}\left( {0,{\sigma^2}} \right)$ is the receiver noise.
This received signal can be utilized to estimate (or predict) the channel realization at any time instant of the block, but the quality of the estimate will reduce with an increase in the delay between the pilot transmission and the considered channel realization. Therefore, we consider the estimates at the channels at time instant $\tau_p+1$, and then these estimates are used as the initial states to get estimates of the channels at all other time instants.
To simplify the notation, we define $\lambda  = {\tau _p} + 1$.
The effective channel at the $t_i$th time instant can be expressed in terms of the channel at the $\lambda$th time instant as
\begin{align} \label{eq:relation-different-instants}
{h_{il}}\left[ t_i \right]={\rho _i}\left[ {\lambda -t_i} \right]{h_{il}}\left[ {\lambda } \right]+{{\bar \rho }_i}\left[ {\lambda -t_i} \right]{\zeta_{il}}\left[ t_i \right],
\end{align}
where ${\zeta_{il}}\left[ {t_i } \right] \sim \mathcal{C}\mathcal{N}\left( {0,{\beta _{il}}} \right)$ denotes the independent innovation component that  relate ${h_{il}}\!\left[ t_i \right]$ and ${h_{il}}\!\left[ {\lambda} \right]$. Using \eqref{eq:relation-different-instants}, we rewrite \eqref{zl} as
\begin{align}\label{z_l}
{z_l}\left[ t_k \right] &=\sqrt {{p_k}}{\rho _k}\left[ {\lambda  - t_k} \right]{h_{kl}}\left[ {\lambda } \right] + \!\!\!\!\sum\limits_{{i \in {\mathcal{P}_k} / \{ k \} }} \!\!\!{\sqrt {{p_i}} {\rho _i}\left[ {\lambda  - t_i} \right]{h_{il}}\left[ {\lambda } \right]}  \notag\\
&+ \sum\limits_{i \in {\mathcal{P}_k}} {\sqrt {{p_i}} {{\bar \rho }_i}\left[ {\lambda  - t_i} \right]{\zeta_{il}}\left[ t_i \right]}  + {w_l}\left[ t_k \right].
\end{align}
Using standard minimum mean square error (MMSE) estimation \cite{bjornson2017massive}, each AP $l$ can compute the MMSE estimate ${{\hat h}_{kl}}\left[ {\lambda } \right]$ of the channel coefficient ${h_{kl}}\left[ {\lambda } \right]$ as
\begin{align}\label{hhat}
{{\hat h}_{kl}}\left[ {\lambda } \right]= \frac{{{\rho _k}\left[ {\lambda  - {t_k}} \right]\sqrt {{p_k}} {\beta _{kl}}}}{{\sum\limits_{i \in {\mathcal{P}_k}} {{p_i}{\beta _{il}}}  + {\sigma^2}}}{z_l}\left[ t_k \right].
\end{align}
The estimate ${{\hat h}_{kl}}\left[ {\lambda } \right]$ and the estimation error ${{\tilde h}_{kl}}\left[ {\lambda } \right]=h_{kl}\left[ {\lambda } \right]-{{\hat h}_{kl}}\left[ {\lambda } \right]$ are distributed as ${\mathcal{CN}}\left( {0, {\gamma _{kl}} } \right)$ and ${\mathcal{CN}}\left( {0,{\beta _{kl}} - {\gamma _{kl}} } \right)$, respectively,
where
\begin{align} \label{eq:variance-estimate}
{\gamma _{kl}}  \triangleq \frac{{\rho _k^2\left[ {\lambda  - {t_k}} \right]{p_k}\beta _{kl}^2}}{{\sum\limits_{i \in {\mathcal{P}_k}} {{p_i}{\beta _{il}}}  + {\sigma^2}}}.
\end{align}
\begin{rem}
The channel estimate ${{\hat h}_{kl}}\left[ {\lambda } \right]$ is degraded by the signals transmitted by the pilot-sharing UEs at the time instant $t_k$. This represents the pilot contamination effect \cite{Ngo2017Cell}.
\end{rem}

\section{Performance Analysis}\label{se:performance}
In this section, we investigate the uplink performance of CF massive MIMO systems with channel aging within each resource block. Meanwhile, we consider small-cell systems for comparison. Novel SE expressions are derived for both systems. In addition, a practical FPC is applied in both systems to further improve the system performance.

\newcounter{mytempeqncnt1}
\begin{figure*}[t!]
\normalsize
\setcounter{mytempeqncnt}{1}
\setcounter{equation}{16}
\begin{align}\label{LSFD}
{\mathrm{SE}}_k^{{\mathrm{LSFD}}} = \frac{1}{{{\tau _c}}}\sum\limits_{n = \lambda }^{{\tau _c}} {{{\log }_2}\left( {1 + \rho _k^2\left[ {n - \lambda } \right]{p_k}{\mathbf{b}}_k^{\mathrm{H}}{{\left( {\sum\limits_{i = 1}^K {{p_i}{{\mathbf{\Gamma }}_{ki}}}  + \sum\limits_{i \in {\mathcal{P}_k}\setminus \{ k \}} {\rho _i^2\left[ {n - \lambda } \right]{p_i}{{\mathbf{c}}_{ki}}{\mathbf{c}}_{ki}^{\mathrm{H}}}  + {\sigma^2}{{\mathbf{\Lambda }}_k}} \right)}^{ - 1}}{{\mathbf{b}}_k}} \right)}.
\end{align}
\setcounter{equation}{7}
\hrulefill
\end{figure*}

\subsection{CF Massive MIMO Systems}

Each AP makes a local estimate of the uplink data using its local channel estimates. These data estimates are then sent to the CPU for joint data detection.
During the uplink data transmission, the received complex baseband signal $y_l[n]$ at AP $l$ during the instants $\lambda \leqslant n \leqslant {\tau _c}$  is given by
\begin{align}\label{y_l}
{y_l}\left[ {n } \right] = \sum\limits_{i = 1}^K {{h_{il}}\left[ {n } \right]} {s_i}\left[ {n } \right] + {w_l}\left[ {n } \right],
\end{align}
where $s_{i}\left[ {n } \right] \sim \mathcal{CN}\left(0, p_{i}\right)$ is the transmit signal from UE $i$ with power $p_i$, and $w_{l} \left[ {n } \right]$ is the receiver noise. In addition, ${h_{il}}\left[ n \right]$ can be expressed in terms of $h_{il} \left[\lambda  \right]$ as
\begin{align}\label{h_il}
{h_{il}}\left[ n \right] = {\rho _i}\left[ {n - \lambda } \right]{h_{il}} \left[\lambda  \right] + {{\bar \rho }_i}\left[ {n - \lambda } \right]{\vartheta_{il}}\left[ n \right],
\end{align}
where $h_{il}\left[\lambda  \right]$ is the initial state in the uplink data transmission phase. The MMSE estimate ${{\hat h}_{il}}\left[\lambda  \right] \sim{\mathcal{CN}}\left( {0, {\gamma _{il}} } \right)$ of $h_{il}\left[\lambda  \right]$ was derived in \eqref{hhat}. Moreover, ${\vartheta_{il}}\left[ {n } \right] \sim \mathcal{C}\mathcal{N}\left( {0,{\beta _{il}}} \right)$ denotes the independent innovation component relating ${h_{il}}\left[ n \right]$ and ${h_{il}} \left[\lambda  \right]$.

To detect the symbol transmitted from the $k$th UE, the $l$th AP multiplies the received signal ${y_l}\left[ {n } \right]$ with the conjugate of its (locally obtained) channel estimate.
Then the obtained quantity ${{\overset{\lower0.5em\hbox{$\smash{\scriptscriptstyle\smile}$}}{s} }_{kl}}\left[ {n } \right] \triangleq \hat h_{kl}^* \left[\lambda  \right] {y_l}\left[ {n } \right]$ is sent to the CPU via the fronthaul. The CPU uses the weights $a_{kl} [n]$ to obtain ${{\hat s}_k}\left[ {n} \right]$ as shown in \eqref{uatf} at the top of the page, where $\mathrm{DS}_k$ represents the desired signal, $\mathrm{BU}_k$ represents the beamforming gain uncertainty, $\mathrm{CA}_k$ represents the channel aging effect,  $\mathrm{UI}_{ki}$ represents the interference caused by transmitted data from other UEs, and $\mathrm{NS}_k$ represents the noise term, respectively.

\begin{rem}
The channel aging effect clearly degrades both the desired signal and beamforming gain uncertainty since $\rho_k^2{\left[ {n - \lambda } \right]}$ reduces as $n$ increases. Focusing on the channel aging quantities, the SINR can be written approximately as $\mathrm{SINR}_k\left[ {n} \right]\!\approx\! 1/\left( {a \kappa \! +\! b } \right)$, where $\kappa  \!\triangleq\! 1/{\rho_k^2{\left[ {n \!-\! \lambda } \right]}}$ and $a$, $b$ are constants depending on the transmit power, channel information, and receiver noise. It is clear that the SINR significantly reduces as the channel aging effect becomes large.
\end{rem}
\begin{thm}\label{th1}
A lower bound on the capacity of UE $k$ is expressed in \eqref{SINR_k} at the top of this page, where
\setcounter{equation}{11}
\begin{align}
{{\mathbf{a}}_k} [n] &\triangleq {\left[ {{a_{k1}} [n]\ldots {a_{kL}}} [n]\right]^{\mathrm{T}}} \in {\mathbb{C}^L}, \\
 {{\mathbf{b}}_k} &\triangleq {\left[ {{\gamma _{k1}} \ldots {\gamma _{kL}}} \right]^{\mathrm{T}}} \in {\mathbb{C}^L}, \hfill \\
  {{\mathbf{\Gamma }}_{ki}} &\triangleq {\mathrm{diag}}\left( {{\gamma _{k1}}  {\beta _{i1}}, \ldots ,{\gamma _{kL}} {\beta _{iL}}} \right) \in {\mathbb{C}^{L \times L}}, \hfill \\
  {{\mathbf{c}}_{ki}} &\triangleq {\left[ {\sqrt {{\gamma _{k1}}  {\gamma _{i1}}  }  \ldots \sqrt {{\gamma _{kL}}{\gamma _{iL}}} } \right]^{\mathrm{T}}} \in {\mathbb{C}^L} \\
  {{\mathbf{\Lambda }}_k} &\triangleq {\mathrm{diag}}\left( {{\gamma _{k1}} , \ldots ,{\gamma _{kL}}} \right) \in {\mathbb{C}^{L \times L}}.
\end{align}
\end{thm}
\begin{IEEEproof}
Please refer to Appendix A.
\end{IEEEproof}

We will call the lower bound on the capacity in Theorem~\ref{th1} an achievable SE. It can be achieved by using a set of $\tau_c-\tau_p$ channel codes for AWGN channels, each spanning over the signal transmitted at the $n$th time instant in every resource block, for $n=\lambda,\ldots,\tau_c$.
The weight vector ${{\mathbf{a}}_k} [n]$ can be different for each $n$ and can be optimized by the CPU to maximize the SE, utilizing the LSFD receiver cooperation approach from \cite{nayebi2017precoding,bjornson2019making}.

\begin{cor}
The effective SINR of UE $k$ is maximized by
\begin{align}
\!\!{{\mathbf{a}}_k} [n]\!= \!\!{\left( {\sum\limits_{i = 1}^K {{p_i}{{\mathbf{\Gamma }}_{ki}}}  \!+ \!\!\!\sum\limits_{i \in {\mathcal{P}_k}\setminus \{ k \}}\!\!\!{\rho _i^2{\left[ {n\!-\lambda } \right]}{p_i}{{\mathbf{c}}_{ki}}{\mathbf{c}}_{ki}^{\mathrm{H}}}  \!+ \!{\sigma^2}{{\mathbf{\Lambda }}_k}} \!\!\right)^{ \!\!- 1}}\!\!\!{{\mathbf{b}}_k}, \notag
\end{align}
which leads to the maximum SE in \eqref{LSFD} at the top of the page.
\end{cor}

If we want to reduce the complexity of LSFD, then the conventional MF receiver cooperation from \cite{Ngo2017Cell} is obtained by using equal weights ${{\mathbf{a}}_k} [n] = {\left[ {1/L \ldots 1/L} \right]^{\mathrm{T}}}$.

\newcounter{mytempeqncnt2}
\begin{figure*}[t!]
\normalsize
\setcounter{mytempeqncnt}{1}
\setcounter{equation}{19}
\begin{align}\label{w}
{\mathrm{SE}}_k^{{\mathrm{small-cell}}} = \mathop {\max }\limits_{l \in \left\{ {1, \ldots ,L} \right\}} \frac{1}{{{\tau _c}}}\sum\limits_{n = \lambda }^{{\tau _c}} {\frac{{{e^{\frac{1}{{{w_{kl}}\left[ n \right]\left( {1 + {A_{kl}}\left[ n \right]} \right)}}}}{E_1}\left( {\frac{1}{{{w_{kl}}\left[ n \right]\left( {1 + {A_{kl}}\left[ n \right]} \right)}}} \right) - {e^{\frac{1}{{{w_{kl}}\left[ n \right]{A_{kl}}\left[ n \right]}}}}{E_1}\left( {\frac{1}{{{w_{kl}}\left[ n \right]{A_{kl}}\left[ n \right]}}} \right)}}{{\ln \left( 2 \right)}}}.
\end{align}
\setcounter{equation}{7}
\hrulefill
\end{figure*}


\subsection{Small-Cell Systems}
In this section, we compare with a conventional small-cell system that consists of $L$ APs and $K$ mobile UEs.
The APs and UEs are at the same locations as in the cell-free case but each UE is only served by one AP, namely the one that gives the highest SE \cite{bjornson2019making}.

In the uplink, each AP first estimates the channel based on signals sent from the UEs, as described earlier. The so-obtained channel estimate is used to detect the desired signal.  The received uplink signal at the $l$th AP is
\setcounter{equation}{17}
\begin{align}\label{small cell}
  &{y_{{l}}}\left[ n \right] = \sum\limits_{i = 1}^K {{h_{il}}\left[ n \right]} {s_i}\left[ n \right] + {w_{{l}}}\left[ n \right] \notag \\
   &= {\rho _k}\left[ {n - \lambda } \right]{{\hat h}_{k{l}}}\left[ {\lambda } \right]{s_k}\left[ n \right]+\underbrace {{\rho _k}\left[ {n - \lambda }\right]{{\tilde h}_{k{l}}}\left[ {\lambda } \right]{s_k}\left[ n \right]}_{{{\mathrm{I}}_{kl,n,1}}} \notag \\
   &+ \underbrace {{{\bar \rho }_k}\left[ {n \!- \lambda } \right]{\vartheta_{k{l}}}\left[ n \right]\!{s_k}\left[ n \right]}_{{{\mathrm{I}}_{kl,n,2}}} +  \underbrace {\sum\limits_{i \ne k}^K {{h_{i{l}}}\left[ n \right]} {s_i}\left[ n \right]}_{{{\mathrm{I}}_{kl,n,3}}}\!+ {w_{{l}}}\left[ n \right],
\end{align}
where the first term denotes the desired received signal from UE $k$, ${w_{{l}}}\left[ {n} \right] \sim \mathcal{C}\mathcal{N}\left( {0,{\sigma^2}} \right)$ is the receiver noise at AP $l$. The remaining terms ${{{\mathrm{I}}_{kln,1}}}$, ${{{\mathrm{I}}_{kln,2}}}$ and ${{{\mathrm{I}}_{kln,3}}}$ are uncorrelated and represent interference caused by channel estimation errors, the channel aging effect, and data transmitted from other UEs. We can treat these as an effective noise $\partial \left[ n \right]={{{\mathrm{I}}_{kl,n,1}}}+{{{\mathrm{I}}_{kl,n,2}}}+{{{\mathrm{I}}_{kl,n,3}}}$. By utilizing \cite[Cor.~1.3]{bjornson2017massive}, we can obtain an achievable SE of UE $k$ at the $n$th time instant as
\begin{align}\label{SE_kl}
{\mathrm{S}}{{\mathrm{E}}_{kl}^{{\mathrm{small-cell}}}}\!\left[ n \right] \!= \!\mathbb{E}\!\left\{ \!{{{\log }_2}\!\left(\! {1 \!+\! \frac{{\rho _k^2\left[ {n - \lambda } \right]{p_k}{{\left| {{{\hat h}_{kl}}\left[ {\lambda } \right]} \right|}^2}}}{{\mathbb{E}\!\left\{ {{{\left| {\partial \left[ n \right]} \right|}^2}\left| {{{\hat h}_{kl}}\left[ {\lambda } \right]} \right.} \right\} \!+ \!{\sigma^2}}}} \!\right)} \!\!\right\}.
\end{align}
\begin{thm}
The achievable SE of UE $k$ can be expressed in closed form as \eqref{w} at the top of this page, where
\setcounter{equation}{20}
\begin{align}
{A_{kl}}\left[ n \right] &= \sum\limits_{i \in {\mathcal{P}_k} \setminus \{ k \} }^K {{{\left( {\frac{{{\rho _i}\left[ {n - \lambda } \right]{\rho _i}\left[ {\lambda  - {t_i}} \right]{p_i}{\beta _{il}}}}{{{\rho _k}\left[ {n -\lambda } \right]{\rho _k}\left[ {\lambda  - {t_k}} \right]{p_k}{\beta _{kl}}}}} \right)}^2}}, \\
{w_{kl}}\left[ n \right] &= \frac{{\rho _k^2\left[ {n - \lambda } \right]{p_k}{\gamma _{kl}}}}{{\sum\limits_{i = 1}^K {{p_i}{\beta _{il}}}  - \sum\limits_{i \in {\mathcal{P}_k}}^K {\rho _i^2\left[ {n - \lambda } \right]{p_i}{\gamma _{il}}}  + {\sigma^2}}},
\end{align}
and ${E_1}\left( x \right) = \int_1^\infty  {\frac{{{e^{ - xu}}}}{u}} du$ denotes the exponential integral. The UE is served by the AP whose index $l$ is maximizing the expression in \eqref{w}.
\end{thm}
\begin{IEEEproof}
Please refer to Appendix B.
\end{IEEEproof}

\subsection{Power control}

The uplink transmit powers must be selected to mitigate near-far effects.
We will extend the FPC policy presented in \cite{nikbakht2019uplink} to the considered system. In the small-cell system, the transmit power of UE $k$ in the cell $l_k$ is selected as
\begin{align}\label{FPC_SC}
{p_k} = \frac{{{\beta _{k{l_k}}}}}{{\sum\nolimits_{i = 1}^K {{\beta _{i{l_i}}}} }}{P_{\text{sum} }},
\end{align}
where ${P_{\text{sum} }}$ denotes the sum of power for all UEs.
FPC in CF massive MIMO systems depends on all the large-scale fading coefficient that involve a given UE, reflecting the effective connection between the UE and all the APs, where the transmit signal power of UE $k$ is
\begin{align}\label{FPC_CF}
{p_k} = \frac{{\sum\nolimits_{l = 1}^L {{\beta _{kl}}} }}{{\sum\nolimits_{i = 1}^K {\sum\nolimits_{l = 1}^L {{\beta _{il}}} } }}{P_{\text{sum} }}.
\end{align}


\section{Numerical Results and Discussion}\label{se:numerical}
We consider a simulation setup where $L=100$ APs and $K=20$ UEs are independently and uniformly distributed within a square of size $0.5 \ {\mathrm{km}}\times0.5\ {\mathrm{km}}$. We utilize the three-slope propagation model from \cite{Ngo2017Cell} and
consider communication at the carrier frequency $f_c=2$ GHz. All UEs has the maximum power $P_{\max}=20$ dBm in full power transmission, the sum of power for all UEs is $P_{\text{sum}}=KP_{\max}$, the bandwidth is $\mathrm{B}=20$ MHz, the noise power is $\sigma^2\!=\!-96$ dBm. The length of resource block is $2$ ms and contains $\tau_c\!=\!200$ samples.

\begin{figure}[t]
\centering \vspace{-3mm}
\includegraphics[scale=0.48]{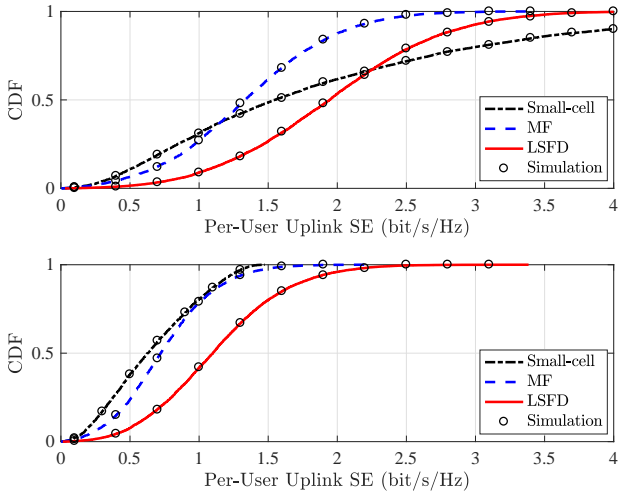}
\caption{The CDF of per-user uplink SE of CF and small-cell systems with full power ($L=100, K=20, \tau_p=10$). (a) $f_DT_s=0$; (b) $f_DT_s=0.002$.}
\label{figure_OP}
\end{figure}

Fig.~\ref{figure_OP} compares the CDF of the per-user uplink SE for CF massive MIMO and small-cell systems with full power with $f_dT_s\!=\!0$ and $0.002$, respectively. The randomness is due to the random AP and UE locations.
It is clear that the LSFD system performs better than MF and small-cell systems at the median and 95\%-likely SE points.
Increasing the normalized Doppler shift $f_DT_s$ from 0 to 0.002 causes $43.5\%$ median SE loss of LSFD, $45.8\%$ median SE loss of MF and $59.3\%$ median SE loss of small-cell, respectively. The reason is that LSFD utilize the knowledge of the fading statistics in the entire network to calculate weight coefficient and thereby mitigate interference, which can effectively counter the impact of channel aging.
Note that the small-cell system makes use of a tighter capacity bound that utilizes the channel estimates in the data detection but \mbox{anyway performs poorly under channel aging.}

\begin{figure}[t] \vspace{-3mm}
\centering
\includegraphics[scale=0.48]{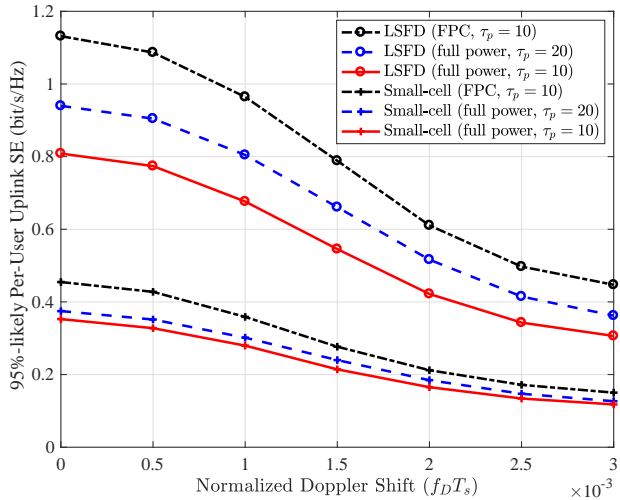}
\caption{95\%-likely per-user uplink SE against the value of $f_DT_s$ for CF and small-cell systems ($L=100$, $K=20$).}
\label{figure_ABEP}
\end{figure}

The 95\%-likely per-user uplink SE with LSFD and of the small-cell system is shown in Fig.~\ref{figure_ABEP}, as a decreasing function of the normalized Doppler shift $f_DT_s$. We notice that CF with LSFD achieves larger 95\%-likely SE than the corresponding small-cell system in both low- and high-mobility conditions. Therefore, CF massive MIMO systems are more suitable for mobility scenarios than small-cell systems.
For both types of systems, the 95\%-likely SE with FPC is getting closer to the 95\%-likely SE with full power when $f_DT_s$ varies from 0 to 0.003, especially for small-cell systems.
The reason is that the self-interference caused by the channel aging effect becomes more dominant than the inter-user interference in high-mobility scenarios, thus the interference reduction due to FPC becomes less influential.
Furthermore, the large-scale fading coefficients available at the CPU can be efficiently combined for the power control in CF massive MIMO systems.
The figure also shows results for the case when the length of the training phase $\tau_p$ is increased, which leads to better SE. 



\section{Conclusions}\label{se:conclusion}
This paper investigates the performance of CF massive MIMO and small-cell systems, taking into account the impact of channel aging and pilot contamination. Based on the derived channel estimates, we derive novel and exact closed-form expressions for the uplink SE of both systems and quantify the channel aging effect. We notice that the channel aging effect degrades the performance of both systems, but the CF massive MIMO systems is less affected by small-cell systems in mobility scenarios. A practical FPC is applied to improve the performance of both systems, but the gain from FPC is gradually reduced as the channel aging becomes stronger.

\begin{appendices}
\section{Proof of Theorem 1}

We consider a set of $\tau_c-\tau_p$ channel codes, each applied to the $n$th time instant in every resource block, for $n=\lambda,\ldots,\tau_c$.
Using the use-and-then-forget capacity bound in \cite{bjornson2017massive} at every time instant and taking the average, the SE of UE $k$ is
\begin{align}
{\text{S}}{{\text{E}}_k} = \frac{1}{{{\tau _c}}}\sum\limits_{n = \lambda }^{{\tau _c}} {{{\log }_2}\left( {1 + {\text{SIN}}{{\text{R}}_k}\left[ n \right]} \right)},
\end{align}
with ${{\text{SIN}}{{\text{R}}_k}\left[ n \right]}$ given as
\begin{align}
\frac{{\mathbb{E}\left\{ {{{\left| {{\text{D}}{{\text{S}}_{k,n}}} \right|}^2}}\right\}}}{{\mathbb{E}\!\left\{ \!{{{\left| {{\text{B}}{{\text{U}}_{k,n}}} \right|}^2}} \!\right\} \!+ \!\mathbb{E}\!\left\{ \!{{{\left| {{\text{C}}{{\text{A}}_{k,n}}} \right|}^2}} \!\right\} \!+ \! \sum\limits_{i \ne k}^K \!{\mathbb{E}\!\left\{ \!{{{\left| {{\text{U}}{{\text{I}}_{ki,n}}} \right|}^2}} \!\right\}} \! + \! \mathbb{E}\!\left\{\! {{{\left| {{\text{N}}{{\text{S}}_{k,n}}} \right|}^2}} \!\right\}}}. \notag
\end{align}
We will compute every term of ${{\text{SIN}}{{\text{R}}_k}\left[ n \right]}$ to obtain \eqref{SINR_k}.

\emph{1) Compute $\mathbb{E}\!\!\left\{\!{{{\left| {{\mathrm{D}}{{\mathrm{S}}_{k,n}}} \right|}^2}}\!\right\}$:}
Based on the properties of MMSE estimation, ${\hat h}_{kl}$ and ${\tilde h}_{kl}$ are independent \cite{bjornson2017massive}. Thus, we have
\begin{align}
   \mathbb{E}\!\!\left\{\!\!{{{\left| {{\mathrm{D}}{{\mathrm{S}}_{k,n}}} \right|}^2}}\!\!\right\}  \!\!&= \!\rho _k^2\!\left[ {n\! -\! \lambda } \right]\!{p_k}\!{\left| {\sum\limits_{l = 1}^L \!{a_{kl}^ *\!\left[ {n } \right] \!\mathbb{E}\!\!\left\{ \!{\hat h_{kl}^*\!\left[ { \lambda } \right]\!\!\left(\! {{{\hat h}_{kl}}\!\left[ { \lambda } \right] \!+ \!{{\tilde h}_{kl}}\!\left[ { \lambda } \right]} \!\right)} \!\!\right\}} } \!\right|^2} \hfill \notag \\
   &= \rho _k^2\left[ {n - \lambda} \right]{p_k}{\left| {\sum\limits_{l = 1}^L {a_{kl}^ *\left[ {n  } \right] {\gamma _{kl}}} } \right|^2} .
\end{align}
\emph{2) Compute ${\mathbb{E}\left\{ {{{\left| {{\mathrm{B}}{{\mathrm{U}}_{k,n}}} \right|}^2}} \right\}}$:}
Because the variance of a sum of independent RVs is equal to the sum of the variances, we have
\begin{align}\label{BU1}
  &\mathbb{E}\left\{ {{{\left| {{\mathrm{B}}{{\mathrm{U}}_{k,n}}} \right|}^2}} \right\} = \rho _k^2\left[ {n-\lambda} \right]{p_k}\notag\\
  &\times \sum\limits_{l = 1}^L {{{\left| {a_{kl}^ *\left[ {n  } \right] } \right|}^2}} {\left( { {\Upsilon _1} - {{\left| {\mathbb{E}\left\{ {\hat h_{kl}^*\left[ {\lambda  } \right]{h_{kl}}\left[ {\lambda  } \right]} \right\}} \right|}^2}} \right)},
\end{align}
where
\begin{align}\label{g1}
{\Upsilon _1}&=\mathbb{E}\left\{ {{{\left| {\hat h_{kl}^*\left[ {\lambda  } \right]{h_{kl}}\left[ {\lambda  } \right]} \right|}^2}} \right\}=\mathbb{E}\left\{ {{{\left| {\hat h_{kl}^*\left[ {\lambda  } \right]{{\tilde h}_{kl}}\left[ {\lambda  } \right]} \right|}^2}} \right\} \notag\\
&+\mathbb{E}\left\{ {{{\left| {{{\hat h}_{kl}}}\left[ {\lambda  } \right] \right|}^4}} \right\} ={{\gamma _{kl}}\left( {{\beta _{kl}} - {\gamma _{kl}}} \right) + 2\gamma _{kl}^2}.
\end{align}
Plugging \eqref{g1} into \eqref{BU1}, we obtain
\begin{align}
\mathbb{E}\left\{ {{{\left| {{\mathrm{B}}{{\mathrm{U}}_{k,n}}} \right|}^2}} \right\} = \rho _k^2\left[ {n - \lambda} \right]{p_k}\sum\limits_{l = 1}^L {{{\left| {a_{kl}^ * \left[ {n  } \right]} \right|}^2}{\gamma _{kl}}{\beta _{kl}}} .
\end{align}
\emph{3) Compute ${\mathbb{E}\left\{ {{{\left| {{\mathrm{C}}{{\mathrm{A}}_{k,n}}} \right|}^2}} \right\}}$:}
Using \cite[Eq. (91)]{ozdogan2019performance}, we have
\begin{align}\label{q1}
&\mathbb{E}\!\left\{ \!{{{\left| {{\mathrm{C}}{{\mathrm{A}}_{k,n}}} \right|}^2}} \!\right\} \!= \!\bar \rho _k^2\left[ {n \!-\! \lambda} \right]{p_k}\mathbb{E}\!\left\{\! {{{\left| {\sum\limits_{l = 1}^L {a_{kl}^ * \!\left[ {n } \right]}{\hat h_{kl}^*\!\left[ {\lambda  } \right]{\vartheta_{kl}}\!\left[ {n} \right]} } \right|}^2}} \!\right\} \notag\\
 &= \bar \rho _k^2\!\left[ {n\!-\!{\lambda}} \right]\!{p_k}\!\!\left( \!{\sum\limits_{l = 1}^L \!{{{\left| {a_{kl}^ *\!\left[ {n  } \right] } \right|}^2}} {{\Upsilon _2}}\!+\!\sum\limits_{l = 1}^L \!{\sum\limits_{m \ne l}^L \!{{a_{kl}}\!\left[ {n  } \right]\!a_{km}^ *\!\left[ {n  } \right] } {{\Upsilon _3}} } } \!\right).
\end{align}
From the definition of channel aging in \eqref{h_il}, ${{\vartheta_{kl}}\left[ {n} \right]}$ is uncorrelated with ${\hat h_{kl}}\left[ {\lambda  } \right]$. Therefore, we have
\begin{align}
   {\Upsilon _2} &= \mathbb{E}\left\{ {{{\left| {\hat h_{kl}^*\left[ {\lambda  } \right]{\vartheta_{kl}}\left[ {n} \right]} \right|}^2}} \right\}={\gamma _{kl}}{\beta _{kl}}, \notag\\
  {\Upsilon _3} &=\mathbb{E}\left\{ {{{\left( {\hat h_{kl}^*\left[ {\lambda  } \right]{\vartheta_{kl}}\left[ {n} \right]} \right)}^*}\left( {\hat h_{km}^*\left[ {\lambda  } \right]{\vartheta_{km}}\left[ {n} \right]} \right)} \right\} \!=0.\notag
\end{align}
Then, substituting above expressions into \eqref{q1}, we obtain
\begin{align}
\mathbb{E}\left\{ {{{\left| {{\mathrm{C}}{{\mathrm{A}}_{k,n}}} \right|}^2}} \right\} = \bar \rho _k^2\left[ {n - \lambda} \right]{p_k}\sum\limits_{l = 1}^L {{{\left| {a_{kl}^ *\left[ {n  } \right] } \right|}^2}} {\gamma _{kl}}{\beta _{kl}}.
\end{align}
\emph{4) Compute ${\mathbb{E}\left\{ {{{\left| {{\mathrm{U}}{{\mathrm{I}}_{ki,n}}} \right|}^2}} \right\}}$:}
When ${i \in {\mathcal{P}_k} \setminus \{ k \} }$, ${{h_{il}}\left[ {n} \right]}$ is correlated with ${\hat h_{kl}}\left[ {\lambda  } \right]$. Then, we have
\begin{align}\label{t1}
&\mathbb{E}\left\{ {{{\left| {{\mathrm{U}}{{\mathrm{I}}_{ki,n}}} \right|}^2}} \right\} = {p_i}\mathbb{E}\left\{ {{{\left| {\sum\limits_{l = 1}^L {{a_{kl}}\left[ {n  } \right]\hat h_{kl}^*\left[ {\lambda  } \right]{h_{il}}\left[ {n} \right]} } \right|}^2}} \right\} \notag\\
&= \sum\limits_{l = 1}^L {{{\left| {a_{kl}^ *\left[ {n  } \right] } \right|}^2}} {{\Upsilon _4}}  + \sum\limits_{l = 1}^L {\sum\limits_{m \ne l}^L {{a_{kl}}\left[ {n  } \right]a_{km}^ *\left[ {n  } \right] } {{\Upsilon _5}} }.
\end{align}
By utilizing \eqref{h_il}, we obtain
\begin{align}
 \label{s1} &{\Upsilon _4} = \mathbb{E}\left\{ {{{\left| {\hat h_{kl}^*\left[ {\lambda  } \right]{h_{il}}\left[ {n} \right]} \right|}^2}} \right\} \notag\\
  &=  \rho _i^2\left[ {n - \lambda} \right]\mathbb{E}\left\{ {{{\left| {\hat h_{kl}^*\left[ {\lambda  } \right]{h_{il}}\left[ {\lambda  } \right]} \right|}^2}} \right\} + \bar \rho _i^2\left[ {n - \lambda} \right]{\gamma _{kl}}{\beta _{il}}, \\
 \label{s2} &{\Upsilon _5} = \mathbb{E}\left\{ {{{\left( {\hat h_{kl}^*\left[ {\lambda  } \right]{h_{il}}\left[ {n} \right]} \right)}^*}\left( {\hat h_{km}^*\left[ {\lambda  } \right]{h_{im}}\left[ {n} \right]} \right)} \right\} \notag\\
 &= \rho _i^2\left[ {n - \lambda} \right]\mathbb{E}\left\{ {\hat h_{kl}^*\left[ {\lambda  } \right]{h_{il}}\left[ {\lambda  } \right]} \right\}\mathbb{E}\left\{ {\hat h_{km}^*\left[ {\lambda  } \right]{h_{im}}\left[ {\lambda  } \right]} \right\}.
\end{align}
Then, using \eqref{hhat}, we can compute
\begin{align}
  \label{s3} \mathbb{E}\left\{ {{{\left| {\hat h_{kl}^*\left[ {\lambda  } \right]{h_{il}}\left[ {\lambda  } \right]} \right|}^2}} \right\}
   &= {\gamma _{kl}}{\beta _{il}} + \left\{ {\begin{array}{*{20}{c}}
  {{\gamma _{kl}}{\gamma _{il}},i \in {\mathcal{P}_k}} \\
  {0,i \notin {\mathcal{P}_k}}
\end{array}} \right. ,\\
  \label{s4} \mathbb{E}\left\{ {\hat h_{kl}^*\left[ {\lambda  } \right]{h_{il}}\left[ {\lambda  } \right]} \right\}
   &= \left\{ {\begin{array}{*{20}{c}}
  {{\sqrt {{\gamma _{kl}}{\gamma _{il}}} },i \in {\mathcal{P}_k}} \\
  {0,i \notin {\mathcal{P}_k}}
\end{array}} \right. .
\end{align}
Finally, plugging \eqref{s1}, \eqref{s2}, \eqref{s3}, and \eqref{s4} into \eqref{t1}, we obtain
\begin{align}
&\mathbb{E}\left\{ {{{\left| {{\mathrm{U}}{{\mathrm{I}}_{ki,n}}} \right|}^2}} \right\} = {p_i}\sum\limits_{l = 1}^L {{{\left| {a_{kl}^ * \left[ {n  } \right]} \right|}^2}} {\gamma _{kl}}{\beta _{il}} \notag\\
&+ \left\{ {\begin{array}{*{20}{c}}
  {{\rho _i^2}\left[ {n - \lambda} \right]{p_i}{{\left| {\sum\limits_{l = 1}^L {a_{kl}^ *\left[ {\lambda  } \right] \sqrt {{\gamma _{kl}}{\gamma _{il}}} } } \right|}^2},i \in {\mathcal{P}_k}} \\
  {0,i \notin {\mathcal{P}_k}}
\end{array}} \right.
\end{align}
and this finishes the proof.
\section{Proof of Theorem 2}
For all ${i \notin {\mathcal{P}_k}}$, ${{{\hat h}_{il}}\left[ {\lambda} \right]}$ and ${{{\hat h}_{kl}}\left[ {\lambda} \right]}$ are independent. Moreover, for all ${i \in {\mathcal{P}_k}}$, we have
\begin{align}
{{\hat h}_{il}}\left[ {\lambda} \right] = \frac{{{\rho _i}\left[ {\lambda - {t_i}} \right]\sqrt {{p_i}} {\beta _{il}}}}{{{\rho _k}\left[ {\lambda - {t_k}} \right]\sqrt {{p_k}} {\beta _{kl}}}}{{\hat h}_{kl}}\left[ {\lambda} \right].
\end{align}
Using these results, we obtain
\begin{align}\label{partial}
&\mathbb{E}\left\{ {{{\left| {\partial \left[ n \right]} \right|}^2}\left| {{{\hat h}_{kl}}\left[ {\lambda} \right]} \right.} \right\} = \sum\limits_{i \in {\mathcal{P}_k}\setminus \{ k \}}^K {\frac{{\rho _i^2\left[ {n - \lambda} \right]\rho _i^2\left[ {\lambda - {t_i}} \right]p_i^2\beta _{il}^2}}{{\rho _k^2\left[ {\lambda - {t_k}} \right]{p_k}\beta _{kl}^2}}} \notag\\
&\times {\left| {{{\hat h}_{kl}}\left[ {\lambda} \right]} \right|^2} + \sum\limits_{i = 1}^K {{p_i}{\beta _{il}}}  - \sum\limits_{i \in {\mathcal{P}_k}}^K {\rho _i^2\left[ {n -\lambda} \right]{p_i}{\gamma _{il}}}.
\end{align}
Substituting \eqref{partial} into \eqref{SE_kl}, we can expand the expression as
\begin{align}
  &\mathbb{E}\!\left\{ \!{{{\log }_2}\!\left( \!{1 \!+ \!{{\left| {{{\hat h}_{kl}}\left[ {\lambda} \right]} \right|}^2}\frac{{\rho _k^2\left[ {n \!- \!\lambda} \right]{p_k}\left( {1 \!+ \!{A_{kl}}\left[ n \right]} \right)}}{{\sum\limits_{i = 1}^K {{p_i}{\beta _{il}}} \! - \!\!\! \sum\limits_{i \in {\mathcal{P}_k}}^K \!\!{\rho _i^2\left[ {n \!-\! \lambda} \right]{p_i}{\gamma _{il}}}  \!+ \!{\sigma^2}}}} \!\right)} \!\right\} \notag \\
   &- \!\mathbb{E}\!\left\{ \!{{{\log }_2}\!\left( \!\!{1 \!+ \!{{\left| {{{\hat h}_{kl}}\!\left[ {\lambda} \right]} \right|}^2}\!\!\frac{{\rho _k^2\!\left[ {n \!- \!\lambda} \right]{p_k}{A_{kl}}\!\left[ n \right]}}{{\sum\limits_{i = 1}^K {{p_i}{\beta _{il}}}  \!- \!\!\!\sum\limits_{i \in {\mathcal{P}_k}}^K \!\!{\rho _i^2\!\left[ {n \!-\! \lambda} \right]{p_i}{\gamma _{il}}}  \!+\! {\sigma^2}}}} \!\right)} \!\right\}. \notag
\end{align}
Then, with the help of \cite[Lemma 3]{bjornson2010cooperative}, we can compute each of the expectations to obtain the final expression in \eqref{w}.
\end{appendices}

\bibliographystyle{IEEEtran}
\bibliography{IEEEabrv,Ref}

\end{document}